\documentclass[a4paper,prl,aps,epsfig,twocolumn,showpacs,preprintnumbers,superscriptaddress,float,floatfix,amsmath,amssymb]{revtex4-2}

\usepackage{graphics}
\usepackage{graphicx}
\usepackage{dcolumn}
\usepackage{bm}

\usepackage{pgfplots}
\usetikzlibrary{arrows.meta}
\pgfplotsset{compat=1.16}

\begin{document}

\title{Shape Programming in Entropic Tissues}
\date{\today}

\author{Carlos M. Duque}
\affiliation{Max Planck Institute for Molecular Cell Biology and Genetics (MPI-CBG), 01307 Dresden, Germany}
\affiliation{Center for Systems Biology Dresden (CSBD), 01307 Dresden, Germany}
\author{Carl D. Modes}
\affiliation{Max Planck Institute for Molecular Cell Biology and Genetics (MPI-CBG), 01307 Dresden, Germany}
\affiliation{Center for Systems Biology Dresden (CSBD), 01307 Dresden, Germany}
\affiliation{Cluster of Excellence Physics of Life, TU Dresden, 01069 Dresden, Germany}

\begin{abstract}
Epithelial morphogenesis, a signature problem of tissue biology and tissue mechanics, continues to inspire biologists and physicists alike. Many treatments focus on tissue fluidization, apical/basal ratio changes, or mechanical instabilities. In contrast to these approaches, shape-programmable materials, where the local lengths in the material change in a prescribed way, offer an appealing analogy. In this analogy, certain in-plane collective cell behaviors could also actively alter the local lengths in a tissue and therefore provide the ingredients necessary for shape programming. In this Letter we demonstrate that this is indeed the case for directed, active T1 rearrangements of cells. We determine the required shape programming parameters associated to tissue patches with both fixed numbers of rearrangements and patches at steady state between directed T1 events and counterbalancing randomly oriented ones using a simple free-boundary vertex model approach. Along the way we uncover a surprising connection between tissues with active T1 events and the central limit theorem, and through it, the physics of entropic springs.    
\end{abstract}

\maketitle

The grand problem of how cells form tissue continues to receive great interest from biologists and physicists. One major question that remains unresolved is that of morphogenesis, or, how cells in an initially simply-shaped tissue reliably create more complex tissues, as in organ development \cite{morpho-development,develop-mechanics-nelson,buckling-nelson} or the formation of limb buds \cite{boehm2010role,hopyan2011budding,hopyan2017biophysical}. Meanwhile, recent progress in our understanding of \textit{engineered} shape transitions holds considerable promise for next generation devices and novel actuation modalities. A specifically designed exotic material -- e.g. a liquid crystal solid or a NIPA hydrogel -- can be coupled to an external field, such as temperature, light, or hydration, in order to control a pre-designed shape transition \cite{efrati-science-2007,Kim-science-2012,mostajeran2016encoding,PhysRevE.84.021711,aharoni-pnas-2018,lavrentovich-pnas-2018}. These transitions are instantiated by \textit{spontaneous strain} fields in the material whose principal directions can be pre-programmed to deliver the desired shape.

If the establishment of a spontaneous strain field drives engineered shape-programmability, could morphogenesis also employ such a strategy? It has long been known that cells in a developing epithelial tissue engage in complex, collective in-plane behaviors. Further, owing to remarkable advances in live-imaging microscopy, patches of cells performing mixes of growth, extrusion, death, elongation, division, and neighbor rearrangements have now been described with striking quantitative detail \cite{blanchard2009tissue,guirao-elife-2015,tissueminer-etournay}. We posit that these collective behaviors, many of which are actively driven, provide coarse-grained spontaneous strains that could capture and predict shape outcomes.

We therefore seek a mapping that connects actively driven collective cellular rearrangements to the physics of spontaneous strain-mediated shape programmability. We begin by considering the local patch-shaping effects of fixed numbers of oriented rearrangements. We then investigate whether actively driven \textit{unoriented} T1 events can provide a restoring strain allowing for the recovery of more homogeneous patch shapes, and in so doing discover an unexpected connection to the physics of entropic springs. Finally, we determine the effective shape programmability parameters -- the spontaneous strain and spontaneous Poisson ratio -- for a tissue patch at steady state balancing the competing effects of oriented and unoriented active rearrangements.

Consider the well-known vertex model (VM), where cells in a confluent epithelia are modeled by polygons.  This model is particularly suitable to study the mechanical tissue response to topological changes in the underlying network of cellular junctions. These changes manifest through cellular events such as divisions, extrusions, or neighbor exchanges, often called T1 events \cite{farhadifar2007influence,staple2010mechanics,duclut2022active,DUCLUT2021203746}. T1 events achieve neighbor exchange through the shrinking and disappearance of the junction between neighbors and the subsequent regrowth of a new junction separating different neighbors. They therefore act directly on the neighbor network of the tissue instead of the conventional elastic deformation where individual cells stretch or contract. Such T1 rearrangements, known to be the driver of convergent-extension \cite{WALLINGFORD2002695,wdev.293,10.1242/dev.073007,10.7554/eLife.78787,sknepnek2023generating}, still provide a rich, underappreciated avenue for shape control. 

We fix the cell mechanical properties and assume negligible bond-tension fluctuations. The non-dimensionalized VM work function is given by \cite{farhadifar2007influence,staple2010mechanics,duclut2022active,DUCLUT2021203746}:
\begin{equation}\label{eq:vm}
E=\frac{1}{2}\sum_{\alpha}\left[\left(A^{\alpha}-1\right)^2+\Gamma\left(P^{\alpha}-P_0\right)^2\right]+\frac{\Lambda}{2}\sum_{b\in\partial S}\ell^{b},
\end{equation}
for cells with unit preferred area. Superscripts $\alpha$ and $b$ denote cell and junction identity, respectively, while $\Gamma$ and $\Lambda$ are the perimeter stiffness and line tension on cellular junctions. $\ell^b$ is the junction length. The term $P_0=-\Lambda/2\Gamma$ is a preferred perimeter. The third term, present in tissue patches, $S$, is a line tension on cellular junctions at the patch boundary, $\partial S$. We focus here on the viscolelastic solid regime. We introduce a preferred direction within the tissue by setting a junction angle with increased likelihood of participation in an active T1 event. We call this the \textit{T1 axis}.

In order to describe the changing shape of a tissue patch under actively driven T1 event dynamics, we consider the \textit{deformation pathway} of the patch, which is a discrete sequence of tissue states in which each state is the result of a single random active T1 event and subsequent mechanical relaxation, possibly including ``passive" T1 events that occur naturally during this relaxation. As we are working in the solid regime, such passive T1 events are rare and do not trigger avalanching T1s \cite{Popovic_2021}. For the remainder of this Letter we simply use the term ``T1 event" for an actively driven T1 event. We define $\mathcal{N}_{\text{T1}}$ to be the number of random T1 events composing the deformation pathway and work within a quasi-static approximation where intermediate tissue states reach mechanical stability prior to the next T1 event \cite{farhadifar2007influence,spencer2017vertex}. Additionally, we fix the number of cells in a simulation by dividing a randomly chosen cell whenever a mechanically-driven extrusion occurs \cite{DUCLUT2021203746}. 

In what follows, we study two mechanisms of deformation. Firstly we fix $\mathcal{N}_{\text{T1}}$ and deform the tissue through T1 events that tend to be aligned with an imposed T1 axis. We refer to T1 events of this kind as \textit{directed}. The second mechanism we consider is that of a dynamical steady state of tissue shape reached in the limit of $\mathcal{N}_{\text{T1}}\rightarrow\infty$ together with a counterbalancing restoring force. Surprisingly, the required restoring force can naturally be provided by an entropic mechanism reminiscent of the entropic spring of polymer physics \cite{de1979scaling}.

We now focus on the spontaneous strains generated by the action of many subsequent directed T1 events. We introduce a preferred direction within the tissue by setting a junction angle with increased likelihood of participation in an active T1 event. We call this the \textit{T1 axis}, $\Theta_{\text{T1}}$ and set $\Theta_{\text{T1}}=\pi/2$. We then sample the junctions according to an approximate normal distribution $p\left(\Delta\theta_{b}\right)\sim\exp{\left[-(\Delta\theta_{b})^2/(2\sigma)^2\right]}$, where $\Delta\theta_{b}$ is the angular difference between a junction and $\Theta_{\text{T1}}$. The distribution spread, $\sigma$, which we set to $0.9$, can be used to tune the rate at which a tissue elongates.

\begin{figure}[t]
  \includegraphics[width=1.0\linewidth]{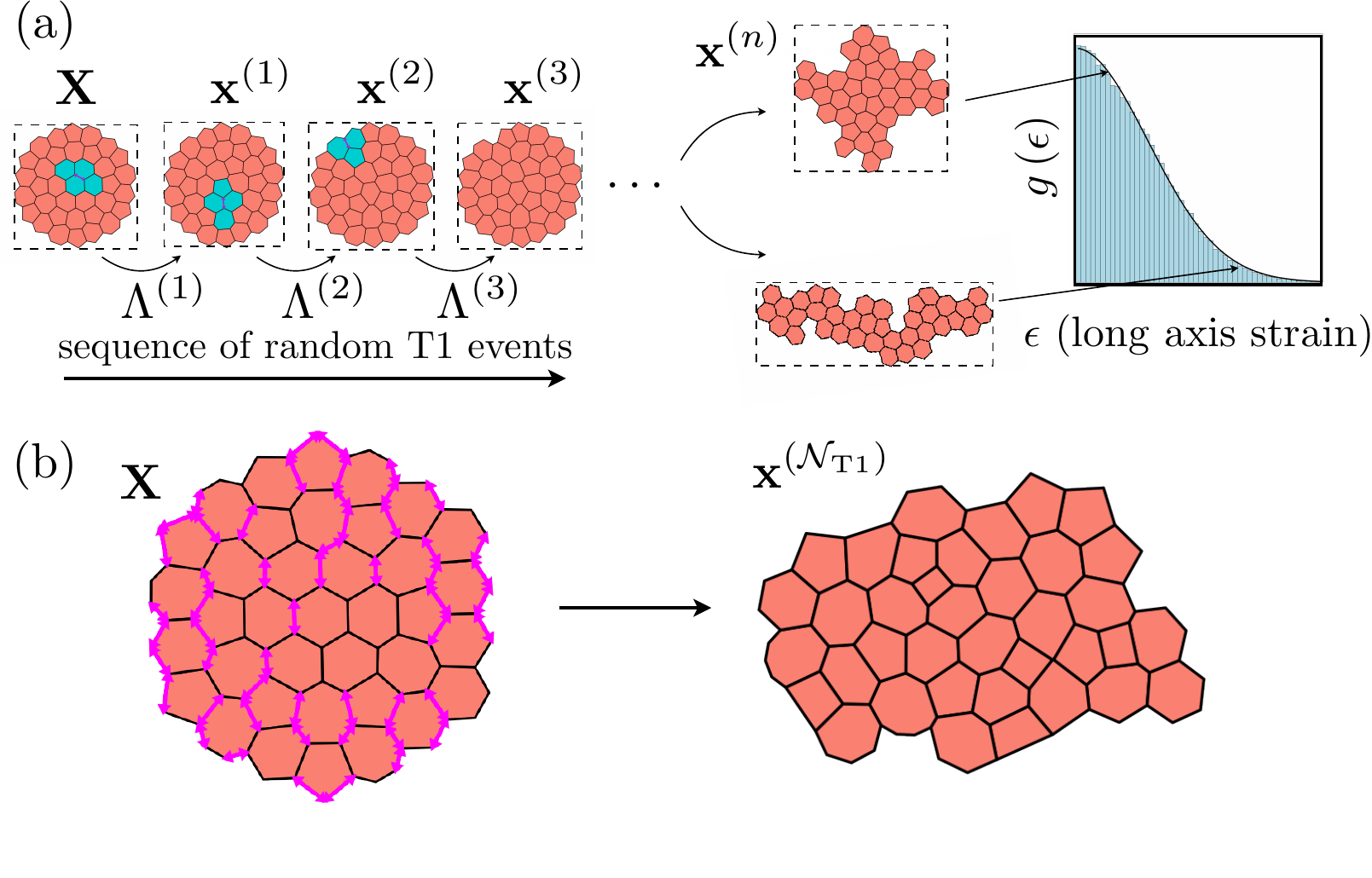}
  \caption{(a) Schematic of Markov-chain type process driven by the subsequent action of random T1 events. The likelihood of each tissue state depends on its long axis strain as shown on the schematic density of states $g(\epsilon)$. (b) An elongated tissue patch resulting from $\mathcal{N}_{\text{T1}}$ simultaneous directed T1 events acting on the starting patch $\mathbf{X}$.}
  \label{fig:fig_chain_histograms}
\end{figure}

Note that different preparations of the tissue are possible. For example, one might apply directed T1s quasi-statically, through a deformation pathway as described previously and depicted in Fig. \ref{fig:fig_chain_histograms}(a). Or, directed T1s may be simultaneously ``pre-loaded" into the tissue and released to act all at once, a scenario depicted in Fig. \ref{fig:fig_chain_histograms}(b). In both cases similar shaping is achieved, though the second case limits the number of directed T1s available. We therefore focus for the remainder of this Letter on quasi-static deformation pathways.

To calculate the spontaneous strain along the tissue's long axis we use the length, $\ell$, of the longer side of the smallest-area rectangle that can enclose the tissue at any given state, Fig. \ref{fig:fig_chain_histograms} (a). $\ell$ is equivalent to the maximum width of the convex hull of the tissue patch \cite{convex_hull}. We start with circular tissue configurations and use the diameter $\ell_0$ to estimate the induced spontaneous strain as $\epsilon=(\ell-\ell_0)/\ell_0$.

\begin{figure}[t]
  \includegraphics[width=1.0\linewidth]{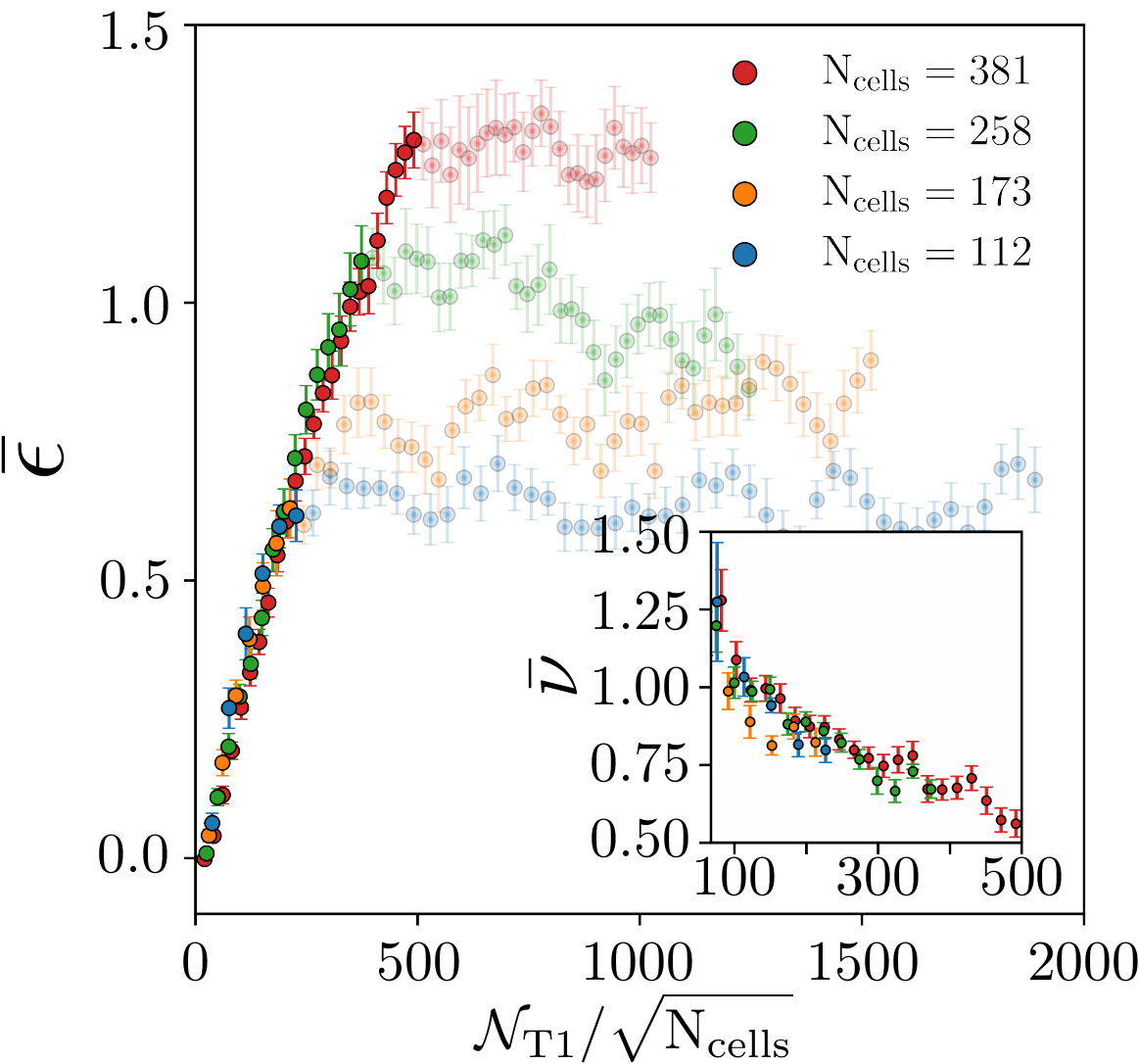}
  \caption{Mean strain, $\bar{\epsilon}$, along the tissue's long axis as a function of $\mathcal{N}_{\text{T1}}/\text{N}_{\text{cells}}^{1/2}$, with VM parameters $\Lambda=0.12$, and $\Gamma=0.04$. The error bars represent the standard error of $\bar{\epsilon}$. A linear fit with slope $\beta\approx0.003$ captures the monotonically increasing regimes. Inset: mean Poisson's ratio, $\bar{\nu}$, as a function of
  $\mathcal{N}_{\text{T1}}/\text{N}_{\text{cells}}^{1/2}$.}
  \label{fig:fixed_t1s}
\end{figure}

In Fig. \ref{fig:fixed_t1s} we show the dependence of the mean spontaneous strain, $\bar{\epsilon}$, on $\mathcal{N}_{\text{T1}}$ demonstrating that topological rearrangements alone can drive a tissue patch into different shape configurations without requiring abrupt deformations of the individual cells. 
The increasing behavior of $\bar{\epsilon}$ with $\mathcal{N}_{\text{T1}}$ underscores the anisotropic reshaping of the tissue.  Note that the values at which $\bar{\epsilon}$ plateaus increase with the system size, a consequence of the fact that patches can adopt more elongated configurations as the number of cells in the tissue increases.

The linear behavior of $\bar{\epsilon}$ in the initial regime highlights the dominant role directed T1 events play in reshaping the patch. In a tissue patch of a given size one expects that individual directed T1 events will, on average, induce the same degree of elongation. Thus, as single, directed T1 elongations are added, the total elongation should scale linearly with $\mathcal{N}_{\text{T1}}$. The relative elongation effect of a single T1 is proportional to the linear size of the patch, yielding the curve collapse evident in Fig. \ref{fig:fixed_t1s}. 

Is it possible that a simple change to the model might allow for dynamical steady states to be reached within the linear elastic regime, before the finite system size-governed stalling of $\bar{\epsilon}$? What is required is an effective restoring force. If the density of states associated to configurations of tissue patches with different elongations is skewed in favor of homogeneous configurations, then an entropic force could naturally provide such a mechanism, similar to that found in an entropic spring \cite{de1979scaling}. Individual T1 events can be thought of as Markov moves acting on the space of patch configurations and therefore, completely randomized, \textit{undirected} T1 events may be capable of providing the effective temperature needed to allow such a restoring force \cite{cugliandolo2011effective}.

Consider the deformation gradient tensor $\Lambda_{ij}=\partial x_i/\partial X_j$, which characterizes changes in infinitesimal distances between a reference elastic configuration, $\mathbf{X}$, and a current configuration, $\mathbf{x}$. Assuming spatially uniform deformations between current and reference states, one may linearly transform the reference state as $x_i=\Lambda_{ij}X_j$. By introducing the displacement vector, $\mathbf{u}$, defined to be $\mathbf{x}-\mathbf{X}$, we may rewrite $\Lambda_{ij}=\delta_{ij}+u_{i,j}$, where $u_{i,j}$ and $\delta_{ij}$ are the gradient of the displacement vector and the Kronecker symbol, respectively. We focus on the case where a current configuration $\mathbf{x}\equiv \mathbf{x}^{(n)}$ can be recast as a series of $n$ previous configurations $\mathbf{x}^{(0)}$, $\mathbf{x}^{(1)}$,..., $\mathbf{x}^{(n-1)}$, with $\mathbf{x}^{(0)}\equiv\mathbf{X}$. Each intermediate state $\mathbf{x}^{(m)}$ is obtained through its previous state by $x_i^{(m)}=\Lambda_{ij}^{(m)}x_j^{(m-1)}$, as shown in Fig. \ref{fig:fig_chain_histograms}(a), implying that our deformation gradient tensors can be decomposed to relate $\mathbf{x}^{(n)}$ with $\mathbf{X}$ as:
\begin{equation}
x_i^{(n)}(\mathbf{X})=\Lambda_{ik}^{(n)}\Lambda_{kl}^{(n-1)}\ldots\,\Lambda_{mj}^{(1)}X_j.
\end{equation}
Defining $\tilde{\Lambda}_{ij}=\Lambda_{ik}^{(n)}\Lambda_{kl}^{(n-1)}\ldots\Lambda_{mj}^{(1)}$ and expressing it in terms of displacement vector gradients we arrive at:
\begin{equation}
\tilde{\Lambda}_{ij}=(\delta_{ik}+u_{i,k}^{(n)})(\delta_{kl}+u_{k,l}^{(n-1)})\ldots(\delta_{mj}+u_{m,j}^{(1)}).
\end{equation}
Provided that the deformations between subsequent intermediate states, $k$, satisfy a small displacement approximation $\left|\partial u_i^{(n-k+1)}(\mathbf{x}^{(n-k)})/\partial x_j^{(n-k)}\right|\ll1$, we keep terms up to linear order in $u_{i,j}$ for all the components of $u_{i,j}$ and write $\tilde{\Lambda}_{ij}=\delta_{ij}+u_{i,j}^{(1)}+u_{i,j}^{(2)}+\ldots+u_{i,j}^{(n)}\equiv\delta_{ij}+\tilde{u}_{i,j}$ \cite{audoly2010elasticity, landau1986theory,chaikin1995principles}, bringing $\tilde{\Lambda}_{ij}$ into the familiar form $\tilde{\Lambda}_{ij}=\delta_{ij}+\tilde{u}_{i,j}$. We can thus use the linear approximation of the strain tensor, $\tilde{\epsilon}_{ij}$, relating reference and current configurations. $\tilde{\epsilon}_{ij}$ then  satisfies the additive property:
\begin{equation}\label{eq:strains}
\tilde{\epsilon}_{ij}=\epsilon_{ij}^{(1)}+\epsilon_{ij}^{(2)}+\ldots+\epsilon_{ij}^{(n)}=\left(\tilde{u}_{i,j}+\tilde{u}_{j,i}\right)/2.   
\end{equation}

\begin{figure}[t]
  \includegraphics[width=1.0\linewidth]{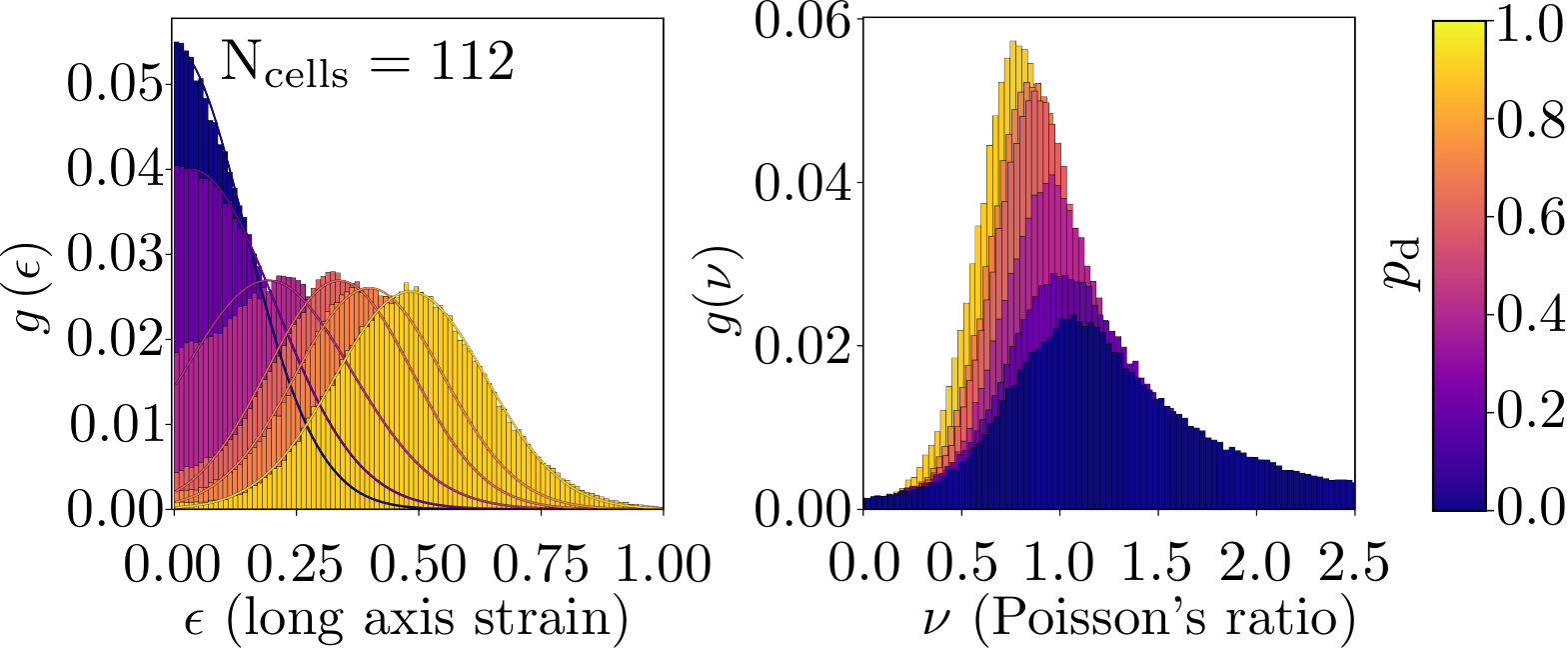}
  \caption{Long axis strain (Poisson's ratio) density of state , $g(\epsilon)$ ($g(\nu)$), for a fixed number of cells $\text{N}_{\text{cells}}$ and increasing directed probabilities $p_{\text{d}}$, all for VM parameters well within the solid regime, $\Lambda=0.12$, and $\Gamma=0.04$. }
  \label{fig:histograms}
\end{figure}

Here, each $\epsilon_{ij}^{(m)}$ is the result of a single T1 event, and as long as the elongation induced by this event is small compared to the long axis of the tissue, we may safely apply the additive strain property. Given that each of the imposed T1 events is of random origin, we assume that the mechanical response of the tissue, as manifested through the additive strains, is also random, even though it may be modulated by the quasi-static approximation that connects subsequent tissue states. The $\epsilon_{ij}^{(k)}$s that \textit{additively} compose $\tilde{\epsilon}_{ij}$ (as per Eq. \ref{eq:strains}) are thus independent and identically distributed random tensors, and we apply the central limit theorem (CLT) to conclude that $\tilde{\epsilon}_{ij}$ is normally distributed \cite{sornette2006critical}.

Now consider a model in which a tissue patch is deformed not only through directed T1 events but also undirected T1 events. As alluded to previously a restoring force must act on the tissue in combination with directed T1s in order to allow for steady states within the elastic regime. The fact that the CLT approximation implies normally distributed net spontaneous strains together with the addition of an effective temperature through the inclusion of undirected T1 events provides the ingredients required for an entropic restoring force. We assume that directed and undirected T1 events occur independently from each other and non-simultaneously. We respectively assign them directed and undirected probabilities $p_{\text{d}}$ and $p_{\text{u}}$ such that $p_{\text{d}}+p_{\text{u}}=1$.

In Fig. \ref{fig:histograms} (a) the density of states with respect to the spontaneous strain, $g(\epsilon)$, is shown for a fixed system size and increasing values of $p_{\text{d}}$. Each density of states was calculated by acting on the tissue patch with a total of $\mathcal{N}_{\text{T1}}=10^5$ events. Each density of states is very well approximated by a Gaussian distribution, supporting the CLT hypothesis. As conjectured, undirected T1 events provide a restoring mechanism, as the peaks of $g(\epsilon)$ shift towards smaller $\epsilon$ as $p_{\text{d}}$ decreases. Moreover, as $p_{\text{d}}\rightarrow0$, the density of states becomes less spread and more peaked, a further testament of the fact that the configuration space of the tissue is dominated by compact rather than ``stringy" configurations. We illustrate this distinction on the late states of the Markov chain schematic of Fig. \ref{fig:fig_chain_histograms}.

Defining $\lambda=\ell/\ell_0$ and $\lambda_{\perp}=\ell_{\perp}/\ell_0$ with $\ell_{\perp}$ the elongation along the short or perpendicular axis of the tissue, a spontaneous version of Poisson's ratio can be calculated as $\nu=-\log{\lambda_{\perp}}/\log{\lambda}$  \cite{PhysRevE.84.021711}. In Fig. \ref{fig:histograms} (b) we show the  Poisson's ratio density of states, $g(\nu)$, for the same parameter values of Fig. \ref{fig:histograms} (a). In contrast to the density of states for $\epsilon$, the distribution of $\nu$ values does not follow the additive property for random strains and the CLT cannot be applied. $\nu$ appears to instead follow a ratio distribution as it can be built by taking the ratio (of the logarithms) of two random variables. Indeed, the $\nu$ density of states are accurately described by skew normal distributions, yielding R squared values close to 1. 

\begin{figure}[t]
  \includegraphics[width=1.0\linewidth]{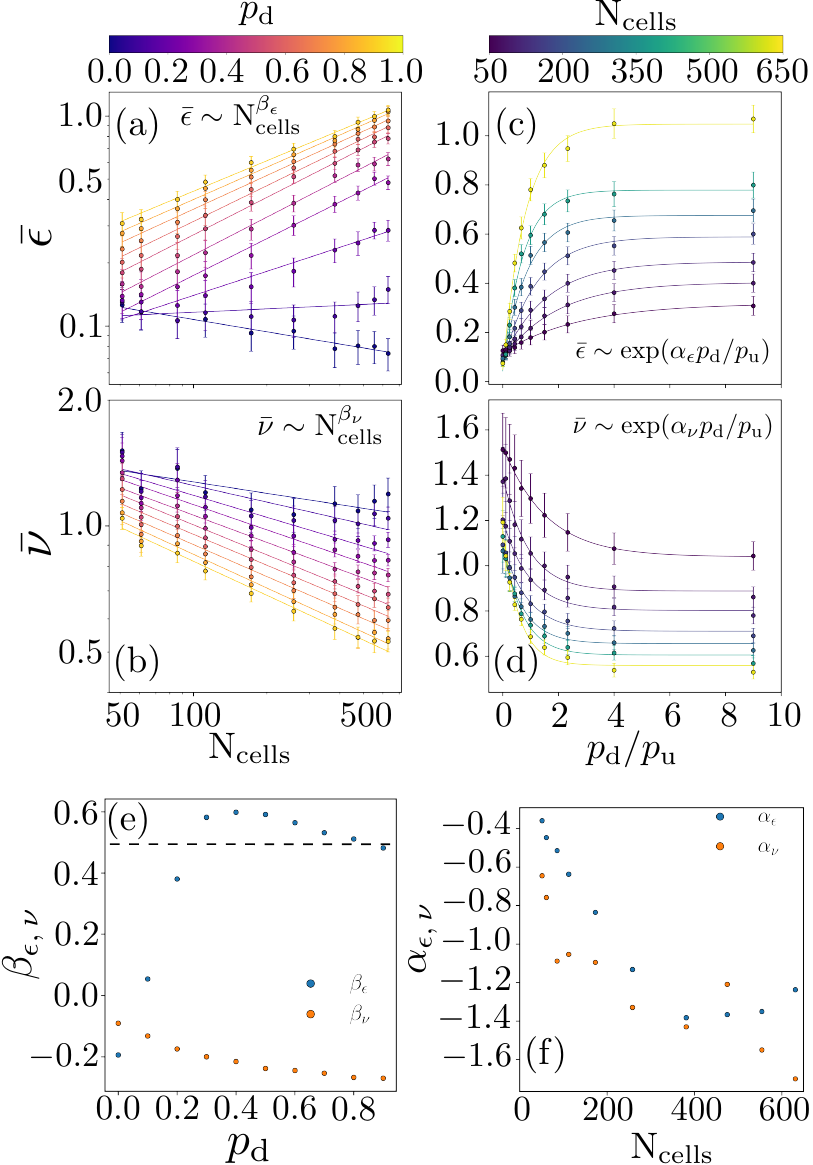}
  \caption{ Shape programmability parameters for varying patch sizes and relative activity, for VM parameters well within the solid regime, $\Lambda=0.12$, and $\Gamma=0.04$. (a-b) Mean long axis strain (Poisson's ratio) as a function of the number of cells $\text{N}_{\text{cells}}$ and increasing directed probabilities $p_{\text{d}}$. (c-d) Mean long axis strain (Poisson's ratio) as a function of $p_{\text{d}}/p_{\text{u}}$ and different number of cells. The error bars in all cases represent the standard errors of the means. (e) Power law exponents $\beta_{\epsilon}$ and $\beta_{\nu}$ for each of the linear fits shown on (a) and (b) respectively. The vertical dashed line at $\beta_{\epsilon}=0.5$ denotes the value at which $\beta_{\epsilon}$ appears to converge as $p_{\text{d}}\rightarrow1$.  (f) Exponential decay constants $\alpha_{\epsilon}$ and $\alpha_{\nu}$ for the exponential fits shown  on (c) and (d) respectively.}
  \label{fig:fig_results}
\end{figure}

We now have the ingredients needed to produce an atlas relating cell intercalations to the parameters of spontaneous strain-mediated shape programmability. In Fig. \ref{fig:fig_results} (a) and (b) we respectively quantify the dependence of the mean spontaneous strain and Poisson's ratio on the system size for increasing fixed values of $p_{\text{d}}$, again indicating the existence of a preferred elongation of the tissue patches as a function of the activity.

Both $\bar{\epsilon}$ and $\bar{\nu}$ are consistent with power-law behavior, at least for the observed decade in $N_{\text{cells}}$. Moreover, we observe two regimes in $p_{\text{d}}$. For $p_{\text{d}}\lesssim0.5$, $\bar{\epsilon}$ monotonically transitions from negative to positive scaling exponents until saturating at values consistent with a power-law scaling of $\frac{1}{2}$ for $p_{\text{d}}\gtrsim0.5$. This is likely the same $\frac{1}{2}$ scaling that collapses the mean strain curves for fixed numbers of directed T1s. For small $p_{\text{d}}$ the entropic shape-restoring force induced by undirected T1s keeps the patch more compact as shown by $\bar{\epsilon}\sim 0$. For $p_{\text{d}}=0$ we observe a clear negative slope implying that shape perturbation from undirected T1 events becomes negligible as the number of cells increases. As $p_{\text{d}}$ increases, directed T1 events bias the strain towards larger values, but the background noise, still dominated by undirected events, keeps the strain relatively small. Only when $p_{\text{d}}\sim p_{\text{u}}$ are there well defined linear trends in $\bar{\epsilon}$. The power-law-like regime of $\bar{\nu}$ is similar. Linearity is not held for $p_{\text{d}}\sim 0$ but is clearly present for increasing $p_{\text{d}}$.

Finally we look at the variation of $\bar{\epsilon}$ and $\bar{\nu}$ with respect to $p_{\text{d}}/p_{\text{u}}$. Here we find that $\bar{\epsilon}$ and $\bar{\nu}$ each exhibit behavior commensurate with exponential decay to an asymptotic value. The decay rates are largely insensitive to $N_{\text{cells}}$, whereas the asymptotic value approached for large $p_{\text{d}}/p_{\text{u}}$ is determined by $N_{\text{cells}}$.

In this Letter, we have generated an explicit mapping between collective cell neighbor rearrangements and the parameters required for active shape programmability in a coarse-grained, continuum model. We have shown these maps both for fixed numbers of directed T1 events as well as dynamical steady states balancing directed T1 events driving elongation and undirected T1 events providing an entropic restoring force. Along the way we have also demonstrated the existence of an unexpected analogy between these tissues and the physics of entropic springs.

Many interesting directions still remain open. Our approach could be generalized to other classes of topological events, including cell division, death, or extrusion. Extending the computational model to larger systems that support gradients in alignment would allow an examination of interactions between the coarse-graining reported here and these gradients, revealing higher order contributions to shape programming outcomes. Furthermore, the status of the steady state tissue as an entropically-driven system could also be expanded on: constructing a formal effective temperature from the activity would allow both explicit stress derivations as well as other applications of traditional thermodynamics. Finally, a more complete exploration of different VM parameter regimes, especially close to the fluid transition, could be useful in capturing varying morphogenetic contexts.

Ultimately, we believe that tissue shape programming through actively established spontaneous strains may provide a new and powerful way to understand the extraordinary shapes and forms of life. Still, how a collective community of cells acquires complex 3D form with robustness and precision remains a deep and beautiful question, and there is much yet to learn.  

\section*{Acknowledgments}
We are happy to acknowledge useful discussions with F. J\"{u}licher, S.W. Grill, M. Popovich, N.A. Dye, A. Materne, J. Fuhrmann, A. Krishna, and M. Staddon. We are particularly grateful to S.W. Grill and M. Staddon for also providing feedback on an in-depth reading of the manuscript. This work was funded by the German Federal Ministry of Education and Research under grant number 031L0160. C.M.D. was further supported by the European Union's Horizon 2020 Research and Innovation Programme under grant agreement no. 829010 (PRIME) during the completion of the manuscript.  

\bibliographystyle{apsrev4-2}
\bibliography{bib}

\appendix

\newpage

\end{document}